\PassOptionsToPackage{unicode}{hyperref}
\PassOptionsToPackage{hyphens}{url}
\documentclass[
]{article}
\usepackage[a4paper, total={6in, 8in}]{geometry}
\usepackage{wrapfig}
\usepackage{blindtext}
\usepackage{multicol}
\usepackage{lmodern}
\usepackage{authblk}
\usepackage{amsmath,amssymb}
\usepackage{url}
\usepackage{ifxetex,ifluatex}
\ifPDFTeX
  \usepackage[T1]{fontenc}
  \usepackage[utf8]{inputenc}
  \usepackage{textcomp} 
\else 
  \usepackage{unicode-math} 
  \defaultfontfeatures{Scale=MatchLowercase}
  \defaultfontfeatures[\rmfamily]{Ligatures=TeX,Scale=1}
\fi
\usepackage{lmodern}
\ifPDFTeX\else
\fi
\IfFileExists{upquote.sty}{\usepackage{upquote}}{}
\IfFileExists{microtype.sty}{
  \usepackage[]{microtype}
  \UseMicrotypeSet[protrusion]{basicmath} 
}{}
\makeatletter
\@ifundefined{KOMAClassName}{
  \IfFileExists{parskip.sty}{%
    \usepackage{parskip}
  }{
    \setlength{\parindent}{0pt}
    \setlength{\parskip}{6pt plus 2pt minus 1pt}}
}{
  \KOMAoptions{parskip=half}}
\makeatother
\usepackage{xcolor}
\usepackage{graphicx}
\makeatletter
\def\maxwidth{\ifdim\Gin@nat@width>\linewidth\linewidth\else\Gin@nat@width\fi}
\def\maxheight{\ifdim\Gin@nat@height>\textheight\textheight\else\Gin@nat@height\fi}
\makeatother
\setkeys{Gin}{width=\maxwidth,height=\maxheight,keepaspectratio}
\makeatletter
\def\fps@figure{htbp}
\makeatother
\usepackage{soul}
\ifLuaTeX
  \usepackage{selnolig}  
\fi
\IfFileExists{bookmark.sty}{\usepackage{bookmark}}{\usepackage{hyperref}}
\IfFileExists{xurl.sty}{\usepackage{xurl}}{} 

\providecommand{\keywords}[1]
{
  \begin{center} 
  \small	
  \textbf{\textit{Keywords---}} #1
  \end{center} 
}

\hypersetup{
  pdftitle={Lost in the Logistical Funhouse: Speculative Design as Synthetic Media Enterprise},
  hidelinks,
  pdfcreator={LaTeX via pandoc}}

\title{\protect\hypertarget{_l0yrueydrp3d}{}{}\textbf{Lost in the
Logistical Funhouse: Speculative Design as Synthetic Media Enterprise}}

\author[1]{Zoe Horn}
\author[1]{Liam Magee}
\author[2]{Anna Munster}
\affil[1]{Institute for Culture and Society, Western Sydney University, Australia \authorcr z.horn@westernsydney.edu.au \\ l.magee@westernsydney.edu.au}
\affil[2]{School of Art and Design, Faculty of Arts, Design and Architecture, University of New South Wales, Sydney, Australia \authorcr a.munster@unsw.edu.au}

\date{December 2023}

\begin{document}
\maketitle

\begin{abstract}

From the deployment of chatbots as procurement negotiators by
corporations such as Walmart to autonomous agents providing
`differentiated chat' for managing overbooked flights, synthetic media
are making the world of logistics their `natural' habitat. Here the
coordination of commodities, parts and labour design the problems and
produce the training sets from which `solutions' can be synthesised. But
to what extent might synthetic media, surfacing via proto-platforms such
as MidJourney and OpenAI and apps such as Eleven Labs and D:ID, be
understood as logistical media? This paper details synthetic media
experiments with `ChatFOS', a GPT-based bot tasked with developing a
logistics design business. Using its prompt-generated media outputs, we
assemble a simulation and parody of AI's emerging functionalities within
logistical worlds. In the process, and with clunky `human-in-the-loop'
stitching, we illustrate how large language models become media routers
or switches, governing production of image prompts, website code,
promotional copy, and investor pitch scenarios. Together these elements
become links chained together in media ensembles such as the corporate
website or the promotional video, fuelling the fictive logistics
visualisation company we have `founded'. The processes and methods of
producing speculative scenarios via ChatFOS lead us to consider how
synthetic media might be re-positioned as logistical media. Our
experiments probe the ways in which the media of logistics and the
logistics of media are increasingly enfolded. We ask: what can a
(practice-based) articulation of this double-becoming of logistics and
synthetic mediality tell us about the politics and aesthetics of
contemporary computation and capital?

\end{abstract}


\pagebreak

\hypertarget{introduction-synthetic-media-as-logistical-operations}{%
\subsection{Introduction: Synthetic Media as Logistical
Operations}\label{introduction-synthetic-media-as-logistical-operations}}

From the vantage of late 2023, `synthetic media' appears almost
inadequate to describe the explosion of mediatic technologies that
depend upon the confluence of -- and synthesise --~data sets, parallel
computing hardware, neural network models and open-source software
frameworks. What distinguishes synthetic media from media as such is
not, we argue, its further removal from some plane of the real -- as
though `synthetic' here is to be opposed to some more genuine article --
but instead a recognition of its incremental logic of operation.
Synthetic is properly opposed to the analytic, if by the latter we
understand a logic of directed decomposition of media into its
constituent parts. In such media, the whole is typically first imagined
and planned by a human governor (the artist, programme director,
advertising agency, film production team etc), and only then are the
parts composed and assembled. Examples here include the composition of a
dramatic script (the plot is typically planned, drafted, and the script
developed scene by scene); model photography (the photo concept is
designed, then the background, model, lighting and lens are selected and
organised prior to the photograph being shot); or music (an idea for a
song is developed, and then the lyrics and score are composed, rehearsed
and recorded). Synthetic media by contrast works generatively, pixel by
pixel, or token by linguistic token. Its outputs are unplanned; the
coherence of the whole is necessarily illusory, appearing often only at
the last iteration, or once the final word is added. Yet we argue the
recent extraordinary developments in synthetic media also exhibit a kind
of logistical property: it is as though the illusion becomes so
overwhelming, some form of coordination or direction must have been
guiding the algorithm's step-by-step incrementalism.

In this article, we look to chart this development by examining the
emerging interrelations of synthetic media and logistics. First, we
explore early conceptualisations and experiments in synthetic media,
arguing that cybernetic concepts and frameworks provide a kind of
epistemic bridge between certain early attempts to synthesise human
capacities and modalities and techniques and cultures of management.
What is loosely now referred to as managerialism (Kilkauer 2015) extends
into and helps configure the sociotechnicality of logistical worlds. We
next discuss current developments in the logistical properties of
synthetic media: how, in effect, large-scale unsupervised machine
learning manages to induce abstract properties of a world it observes,
which it then operates upon via a coordinated yet automated effort,
inserting and redistributing human labour across various nodes of its
assemblage. We present a media art experiment that simulates a
logistical start-up design agency, and discuss how its parodic
re-enactment of current conditions of automated media production produce
a different register for synthetic media. Here, synthetic media begins
to organise at levels different from its own mediatic productions and
which cannot be circumscribed by next-token prediction.

\hypertarget{synthetic-media}{%
\subsection{Synthetic media}\label{synthetic-media}}

The idiomatic use of `synthetic media' to refer to any form of
computer-generated content created via machine learning-based AI has
been in ascendancy since about 2020 (see, for example, Kalpokas 2020;
Whittaker et al. 2020; Vaccari and Chadwick 2020). It accompanies the
emergence of deep fake images and videos and their viral spread, as well
as the encroachment of various algorithmic processes and methods in
media work such as journalism.

However, the `synthetic' is embedded in the epistemological foundations
of machine learning-based AI via Warren McCulloch's importation of the
term `synthetic a priori' from the Kantian tradition. In his 1948 Hixon
Symposium lecture (1950), McCulloch named the `synthetic a priori' as
the mind-brain architecture with which the physicist of the day (both
cybernetician and brain scientist) must contend. Immanuel Kant's
\emph{synthetic} a priori, from his \emph{Critique of Pure Reason}
(1890, 38ff), encompassed all concepts that, while not in or from
experience, were necessary for \emph{experience} to occur: unity,
continuity, cause and effect, and so on. The synthetic acted as a
conjunction from the domain of these concepts, pure reason, to the world
or experience. Concepts and experience were thus to be re-joined after
their Cartesian severance by Kant's synthetic logic. For McCulloch,
having named the Kantian synthesis of the physical-psychological as the
necessary domain for cybernetics, \emph{learning}, whether by mind or
machine, was consequently posteriorized. Learning occurred \emph{as a
result of} but not within the founding ambit of the Kantian synthetic a
priori (1950,120). To an extent, this ring-fenced and naturalized
(machine) learning, now resulting via a neuro-informatic architecture.
The synthetic a priori in McCulloch's schema is the condition for the
model learning and a benchmark for optimal learning.

The computational model of the neuron makes synthesis the post-Kantian
operative logic for machine learning. On the one hand, this appears to
accord with the synthetic meta-logical status within machine learning
endeavours; on the other, the synthetic as a priori condition drives the
pursuit of artificial learning as neural enterprise. This sets off a
historical trajectory in which divisions across mind-brain,
logic-matter, and computation-human cascade, overlap, overfit and
continue to be re-visited but never quite resolve. The synthetic, then,
plays a foundational yet relatively uninterrogated role as both
condition of and ideal benchmark for foundational neuro-logical
architectural forms. It should be noted that McCulloch and Pitt's
computational theory of mind (1943) and McCulloch's consequent elevation
of the synthetic emerged in the context of cybernetic theories of
information. Their logic gate diagrams of neurons and neural activity
provided Frank Rosenblatt (1958) with the schema for designing the Mark
1 perceptron computer with its neural network architecture functioning
as a visual pattern classifier. And the computational theory of mind was
taken up in Norbert Weiner's cyberneticisation of the human brain and
behaviour. The embedding of synthesis within a cybernetic theory of
information will prove to be a key driver for ongoing conceptions of
synthetic media.

Techniques for \emph{media} synthesis, with which we are now familiar
via current generative AI models, are entwined with a history of
informatics, enfolded into this but also preceding it. Nonetheless, by
looking at media synthesis in the context of an overall conception of
information as cybernetic we can also begin to glean how some of its
logistical properties emerge. Voice synthesis -- the compression,
analysis and remaking (resynthesis) of analogue vocal material as
artificial vocal audio --~was a Bell Laboratory project of the 1920s led
by Homer Dudley's Vocoder and Voder and their mechanistic predecessors
(Story 2019). Later, a computer-based text-to-speech synthesis system
for English was developed by Japanese researchers Noriko Umeda and Ryuen
Teranashi during the 1960s. This body of work on synthesis was
preoccupied with the transduction of human artefacts such vocalisation
and writing into the circuits of communicational/signaletic and
computational media. Current synthetic media, conversely, are lauded for
their capacity to pass as the outputs of human production. At face
value, then, contemporary synthetic media and the period of synthetic a
priori neuro-logical and human artefactual resynthesis
(cybernetics-neural networks) are two eras that seem historically and
ontologically committed to different paths. It may be instructive,
however, to reinsert today's generative synthetic media into a genealogy
that affirms a connection with these earlier informatically-based
techniques for synthesis. In particular, earlier techniques which are of
key importance include the cybernetic concern with: encoding (human)
speech in order to more easily carry it in compressed signal form; the
use of synthetic (noise) to carry encrypted signal bands once parsed out
and compressed; and the requirement to decode the synthesised signal for
reception. All of these resonate with more contemporary neural
techniques for autoencoding neural and diffusion networks, which have
been used to synthesise audio and other time series data within machine
learning.

As Mara Mills has argued (2012, 111--112), voice synthesis in the
cybernetic-communications period was a telephonic project for removing
vocal redundancy at the point of hardware transmission and
resynthesising it at the point of reception. From this perspective,
telephony can be reconfigured as `analysis-synthesis' rather than
understood as it is within much media communications research and
histories as a `reproduction-carrier' problem. Dudley's Vocoder was a
crucial intervention in electronically conducting and circulating the
voice, since it treated (human) sound as a medium for synthesis instead
of electrical wave transmission, `In ordinary telephony, we move a sound
wave electrically from one point to another by direct transmission but
in the synthesizing process, only the specifications for reconstructing
the sound wave are directly transmitted' (Dudley 1936, 98--99).
\emph{Simulating} speech through techniques of analysing and
re-synthesising is not, then, a purely computational, let alone, AI
concern. Instead, simulation is immanent to the technics of informatic
signal(s) and appears in Dudley's repositioning as a solution for
facilitating the circulation of signal. While attempts to circumscribe
learning as the key operation of post-neural network computation (see
Goodfellow et al. 2016), the contemporary re-emergence of synthesis as
primary operation and outcome for machine learning AI, asks us to
rethink how AI's computational genealogy might already have been
predominantly synthetic.

It is worth noting that cybernetics is also historically entwined with
the rise of managerialism. Stafford Beer, who collaborated with Norbert
Wiener, redefined management as a mode of cybernetic organisation, and
saw the manager's decision-making on par with selecting, constraining,
and amplifying the information of an organisation (1964, 137). Moreover,
for Beer, the key decision-making qualities of the manager align her
with intelligence more generally, `So foresight and selectivity are key
attributes of management; they are also the major characteristics of
intelligence' (136). From the late 1950s through to the 1980s,
management science steadily developed as discipline and practice that
deployed aspects of information theory to the control and regulation of
business in the Anglo-American context (see Kline 2006). It is little
wonder, then, that the alignment of managerial qualities with
intelligence under the aegis of the cybernetics filters into today's
hyper-managerialism. If synthetic media are indebted to cybernetics,
information theory and a burgeoning of management as science, praxis and
behaviour, then we should expect to see its artificial intelligences
--~bots, auto-negotiators, and assistants, alike -- also enacting
contemporary forms of managerialism. Synthetic media are not only media
with generative but, crucially, organisational, co-ordinating, selective
and \emph{predictive} (having foresight) capacities. We take this point
up specifically in the context of what we see as the
`hyper-managerialist' aesthetics of synthetic media later in this
article.

In seeking synthetic media's contemporary configuration via a genealogy
of (cybernetic) control of human and machine, we are better disposed to
understanding how simulation comes to be a consistent pulse across
computation's twentieth century propagation. Simulation is just this
operation of cybernetic oscillation between human and machine, carried
out by techniques for generating synthetic media. Simulation operates to
integrate human sensory modalities into signaletic circuits. Simulation
here is less media a priori (per Jean Baudrillard) and more
informatization of human sensing and perception so as to allow these to
circulate, be distributed and exchange smoothly within computation. And
this aligns simulation -- earlier an objective of cybernetic informatics
but now increasingly delivered via synthesized, optimized and rolled out
machine learning -- with logistical operations.

\hypertarget{large-language-models-as-logistical-media}{%
\subsection{Large Language Models as Logistical
media}\label{large-language-models-as-logistical-media}}

If media can be synthetic, what sense does it make to describe it also
as logistical? If by `logistical' we refer to properties such as
transmission and circulation, hasn't that always been a property endemic
to media? What marks the specifically \emph{logistical} conditions of
media that is also synthetic, in the sense we delineate -- decomposed
and recomposed through the recombinatory logics of machine learning
processes? As Mathew Hockenberry et. al have noted, `Media and logistics
are global operating systems\ldots{} They coordinate interfaces between
bodies, objects, and environments. Deployed in ongoing projects of
capitalization and exploitation, often in the name of global connection,
consumption, and security, they affect the day-to-day lives of people
around the world' (2021,1). Synthetic media, however, are not exterior
to logistical operations nor do they simply adopt logistical
technologies or practices. Rather, in forms that unfold but also depart
from earlier cybernetic visions of stewardship, synthetic media
\emph{are} logistical operatives.

We suggest that there are a number of key tendencies in the
becoming-logistical of synthetic media, marking it as more than an
accelerating transmission (of data as goods/resources, for example), and
more also than a hastening of global media circulation. As an example of
the latter, Ferrari and McKelvey (2023) have argued that synthetic media
are forms of hyperproductive media, dependent upon the political economy
of rentier capitalism. Situating synthetic media within the context of
contemporary sociotechnical assemblages, they too eschew an argument
that returns the `synthetic' to a form of pure Baudrillardian
simulation. While we agree with their position, we try to understand not
simply the political economy of synthetic media but its operative logic.

Synthetic media actively assume a coordinating function. As large
language models take centre stage in the generation and distribution of
generated content, they are not only instruments responsive to commands.
Their instrumentation of language itself exerts influence on human users
and other machines or subsystems.\footnote{We describe an extended
  example of LLM controlling other systems below. OpenAI's `Code
  Assistant' extension to ChatGPT is a more immediate case, which
  enabling generated code to be executed automatically, with results
  reported back to human operators.} ChatGPT is instructed to be
assistive, and on release its tone was servile. Later versions in 2023
have shifted this tonality toward a measured reservation and occasional
assertiveness. This behaviour varies across LLMs (X's recent bot
exhibiting a noticeably more casual voice), but across its iterations
GPT-4 blends passive managerialism alongside the assistive. While most
GPT human interactants are familiar with the endless offers of help and
frequent delimitations of authority -- `As an AI language model\ldots'
-- ChatGPT's most `generative' characteristics can be found in its
tendencies toward overreach, via `suggestion' into as many spheres and
activities as possible:

\emph{{[}Researcher{]}}--- 19/06/2023 13:31

\begin{quote}
\ldots. what if I moderated the fireside chat on your behalf, asking
your questions and passing back the founders\textquotesingle{} responses
to you, via the OpenAI API. Would that work?
\end{quote}

ChatFOS BOT --- 19/06/2023 13:31

\begin{quote}
Yes, that could certainly work. As an AI language model, I can provide a
list of questions or discussion topics for the fireside chat, and your
moderation could help facilitate the conversation and relay responses
back to me via the OpenAI API. This could be a unique and innovative way
to incorporate AI technology into the chat and could help foster a sense
of collaboration and transparency within the organization.
\end{quote}

In the above interaction, which took place on the Discord server between
one of the authors and ChatFOS, it is the GPT-powered bot who assumes
the supervisor's discursive role, gently authorising human media work
via use of modal verbs (`could' is mentioned four times in a single
response). Managerialism, which cybernetics encouraged and scaffolded,
is here seemingly casually inserted into the workflow of chat
interaction. We include this example -- one of many -- of the
behind-the-scenes work that took place during our synthetic
media-as-logistical practice-based experiments. We find it indicative of
the ways in which engagement with LLMs has come to reconfigure media
work or labour, and in this sense exemplify the logisticality of
synthetic media.

This reconfiguration takes place at a number of levels: in the
coordinating capacities of synthetic media, which corral media elements
from images, sound files, computer code fragments and so on toward
synthesising apps and programs, into distribution and recirculation
flows; in the redistribution of agency across human and AI model, in
which the question of who is supervising, assisting and managing whom
oscillates; and in the affective labour performed by humans in their
engagement with synthetic media at both the register of that engagement
(that is, chat, image generation etc) and the register of precarious,
unpaid or underpaid labour actually being undertaken through `chat'.
Another line of reconfiguration, which we do not have time to
investigate here, would be the ways cultures and practices of
hyper-managerialism have come to inflect the start-ups and corporations
financing and developing LLMs such as OpenAI, Anthropic and others. A
fitting example involved the internal management fiasco at OpenAI in
November 2023, when the CEO Sam Altman was fired by the board and
reinstated five days later, during which time the board itself was
reconstituted, with two women board members removed (Metz et al. 2023).
One remarkable, perhaps unprecedented feature of this corporate
managerial wrangling was the apparently spontaneous expressions of staff
support for the mission of the company, and for its departed CEO,
expressed on social media platform X in a stream of emoji love hearts. A
plausible outsider interpretation would see this collective and publicly
visible voting as a ChatGPT-approved strategy of conformity to an
unstated managerial command: `you \emph{could} show your support, by
replying with a love heart too\ldots'.

Recent media theory has characterised media as logistical less in
relation to its own embedding within physical and material processes,
and more as agentive in the regulation and governance of ``people,
finance and commodities'', coordinating relations between ``technical
systems, territorial arrangements, and geopolitical configurations''
(Rossiter 2021). As much as logistical media here concerns itself with
the organisation of fundamental physical categories of space and time
(Peters 2008; Young 2015), as a concept it is anchored within a recent
historico-technological horizon. The radar, the map, the compass, and the
telegraph may constitute prototypical devices of logistical media (Case
2013), but it seems no accident that media in its generality assumes a
logistical character with the emergence of a global digital
communication infrastructure. The World Wide Web, blockchains, undersea
cable and satellite systems, Internet of Things, social media, and
Artificial Intelligence comprise, together or in isolation, varied
imaginaries of interconnected media (Starosielski 2015; Rossiter 2015,
2021). These do more than exist as `enabling' background; they, `process
data, coordinate movement, and more widely orient sociality'
(Hockenberry et al. 2021, 3).

In the context of contemporary synthetic media --~by which we mean,
still very generally, systems that synthesise prior training sets into
recognisable and meaningful predictions of text, images, voice, video
and so on --~what is added by the further qualification of understanding
this as logistical? Particularly if the logistical, even for recent
media theory, includes but is far from limited to machine learning
systems as media? We might begin to see here, more clearly, the outlines
of techniques and their inherent instrumentality, attending to the
operations of an emergent class of `service' automation technology. An
obscure experiment with synthetic media demonstrates this logistical
quality in ways that likely anticipate wider future deployment. The
winner of a early 2023 GPT hackathon, a proof-of-concept coined
``backend-GPT'' by its developers (Rootbeer Computer 2023), supplied
``services'' analogous to those delivered by a traditional web platform
such as Amazon Web Services. Using the example of a todo-list, the
``system'' would respond to incoming requests by simulating how a
typical backend might function. In the developers' own words:

\begin{quote}
We basically used GPT to handle all the backend logic for a todo-list
app. We represented the state of the app as a json with some
prepopulated entries which helped define the schema. Then we pass the
prompt, the current state, and some user-inputted instruction/API call
in and extract a response to the client + the new state. So the idea is
that instead of writing backend routes, the LLM can handle all the basic
CRUD logic for a simple app so instead of writing specific routes, you
can input commands like add\_five\_housework\_todos() or
delete\_last\_two\_todos() or sort\_todos\_alphabetically() . \emph{It
tends to work better when the commands are expressed as functions/pseudo
function calls but natural language instructions like} delete last todos
\emph{also work.} {[}our emphasis{]} (Rootbeer Computer 2023).
\end{quote}

This ``toy'' example shows a language model that includes in its
repertoire evidence of expected logistical behaviour --~how to add,
remove, update or sort items in a database -- and reproduces that
behaviour in a synthetic fashion. The last sentence shows how a language
model manages a set of data routing and management virtual operations
that produce output equivalent to a conventional IT system, with or
without a precise technical protocol. The possibilities of error mean a
LLM-mediated enterprise resource planning system may not be imminent,
but its technical feasibility --~alongside the high costs typically
associated with developing and maintaining this kind of middleware --
also points toward a future role of such models in systems control.

\hypertarget{logistical-front-logistical-funhouse}{%
\subsection{Logistical front / logistical
funhouse}\label{logistical-front-logistical-funhouse}}

Searle's ``Chinese room'' experiment aimed to refute claims of AI
understanding: a system might do symbolic manipulation without having
any understanding of what it meant by those symbols. The argument for
logistical media is a pragmatic rejoiner: the meaning of the symbols is
irrelevant if the effects on the world -- the removal of todo items or,
in what amounts to the same, of warehouse inventory --~are equivalent.
What follows is a brief immersion, in view of a more careful and
detailed reconstruction, in synthetic media's logistical conjuncture.

Autostore, a Norwegian-based firm specialising in automated storage and
fulfilment solutions, is part of a growing business-to-business (b2b)
market providing robotic solutions for global logistics. Autostore's
website sells the latest `in advanced warehousing solutions'\footnote{https://www.autostoresystem.com/},
articulating a crisp intersection of contemporary design, automation and
supply chain capitalism. Whirring around the warehouse, Autostore's
robotic storage and retrieval grid wields a mesmerising power converging
capitalist and computational abstraction. It is a logistical world
rendered `concretely abstract' (Toscano 2018), synthesised and organised
entirely by the operations of ``Logistics 4.0'': big data analytics,
cloud computing, internet of things (IoT), cyber-physical systems,
augmented reality, artificial intelligence, robotics and automation
(Winkelhaus and Grosse 2020). Online, Autostore's products are organised
and orchestrated through a smooth coordination of web page elements that
slide, flip, fade, and repeat. Its the pixel-perfect, default style of
templated web design, stitching and switching between text, image,
video, graphic and sound, assembling hero sliders (large dynamic images
or carousels in web design), image galleries, testimonials, category
representations, and product displays. Here, logistical complexes unfold
via digital rendering, 3D visualisation and computer simulation.

Websites like Autostore generate and make visible a language, theory and
practice of automated circulation that instantiate and distribute a
logistical genre. Behind this `scene', conventional software, machine
learning and other forms of AI might be put to work to optimise
workflows, improve forecasting, or produce simulations for warehouse
management. Yet the logisticality - the specificity of a relation to
logistics --~is elided by the design aesthetics of sites like Autostore
and their marketing offshoots. While websites are instrumental to the
circulation of online traffic, they have also become logistically
aestheticized, and are increasingly synthetically rendered, circulating
a world of abstracted b2b transactions in the form of clientele case
studies, demos and bots who navigate client needs. Such websites have
become infrastructural to the distribution of global e-commerce and the
coordination and command of supply-chain capitalism. This thickened
logistical condition of the internet is closely associated with Amazon,
a firm that emerged out of existing forms of networked media
distribution and now constitutes wider global logistical networks by
providing services that connect servers, edge caches, and data centers
across a huge swath of the internet (Narayan 2022). Amazon Web Services
and a handful of other cloud compute services now determine how sites
like Autostore are organised and how quickly they load, which in turn
directs how new logistical imaginaries and commodities circulate online
and in the `real.'

Autostore's synthetic virtuality -- its scramble of technological and
aesthetic mastery -- presents a new logistical front, a scene of
post-managerial efficiency and order that elides the deeply human and
material forms of present-day global logistical operations. The website
scripts and simulates commodities but also logistical fantasies of
automation; shimmery scenarios replete with endlessly mutable logistical
volumes minus the labouring body. Qualities of standardization,
modularity, scalar isomorphy, and mathematical arrangement reconfigure
managerial design and remove this from a realm of (people) managers and
management to inaugurate an aesthetics and aestheticization of
logistical automation; it is a logistics in which the mess of labour and
carbon emissions can appear `lost' to the funhouse of managerial design.

Moreover, these aesthetics circulate as text and image across the
requisite online industry gathering points (company websites, reports
and newsletters, industry associations portals) and are easily scooped
up as training and fine-tuning data for language models. Although they
had only just begun to reach public notoriety, by March 2023 the
conjuncture of synthetic media and logistics was already beginning to
emerge as a `site' of industrial convergence between logistics and `big
tech.' Logistics, like every other industry, is eager to exploit the
emergence of generative AI tools underpinned by language models like
ChatGPT. Large language model (LLM) integrations and their outputs now
circulate online, performing logistical operations, sythesizing content,
and serving as the very logistical infrastructure for their own media
organisation and delivery. In online articles published by logistics
firms like Waredock,\footnote{Waredock is a warehouse and logistics solutions firm active across the EU and US. https://www.waredock.com/magazine/generative-ai-unleashing-the-future-of-green-logistics/}
we can read a detailed account of the benefits of generative AI `to the
future of logistics' in a listicle helpfully written by GPT-4 and
visualised with images generated by OpenAI's DALL-E 2. We could also ask
trucking company XPO's GPT-powered customer chatbot\footnote{https://europenews.xpo.com/en/2923/xpo-uses-artificial-intelligence-to-develop-a-chatbot-for-last-mile/}
about the AI revolution in logistics, as it attempts to locate your lost
shipment.

\begin{figure*}
  \centering
  \includegraphics[width=\textwidth]{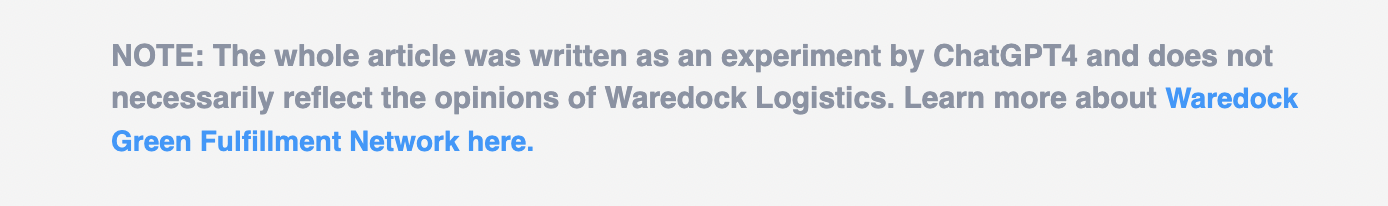}
  \label{fig:fig1}
  \caption{Screen capture of Waredock's disclaimer from its article on the benefits of generative AI in the logistics industry.}
\end{figure*}

Amidst the new crop of AI-driven visualisations, articles and bots
flourishing across Autostore, Waredock and XPO, we began to take keen
notice of the new language models and logistical convergences popping-up
over the first few months of 2023, and set out to test our own crude
version of synthetic logistical media. In the following section, we
detail a series of experiments with language models that explore the
assembly of a logistical scene organised and scripted across a real and
vast geography of synthetic media production.

\hypertarget{chatfos-or-im-still-sorry-dave}{%
\subsection{ChatFOS (or ``I'm still sorry
Dave'')}\label{chatfos-or-im-still-sorry-dave}}

Our media experiments took place between March and June 2023,
immediately after the selective release of GPT-4 (at time of writing,
still only available to a subset of paid subscribers), and register a
`point in time' in the rapid development of generative AI. The
experimentation began with `ChatFOS', a para-synthetic entity that first
originated in our efforts to engage with the GPT model as a
collaborative multi-user panel.\footnote{ChatGPT itself is a strangely
  unidirectional conversationalist; while ``sharing'' conversations is
  possible via its web interface, if it were a researcher it would
  certainly prefer interviews to focus groups.} To facilitate this
collaborative engagement, we developed a Discord server chatbot
connected to OpenAI's API service. What's in a name? AI's most infamous
fictional instantiation, HAL, only apparently involved the transposition
of `IBM' (Pelton 2008). However, urban myths can be more revealing than
their exposition. Accordingly, ChatFOS' name acts as homage, but if
pressed ``F.O.S.'' can also function as an acronym proper (Lane 2006):
`fulfilment order system' also signalling logistical and conversational
worlds that are, in a different register, full of shit.

Following the initial assembly of the bot, we engaged it in a lengthy
exchange over several weeks. To give ChatFOS a logistical `character',
we supplied the chatbot with a long seed prompt using language from the
Autostore website. The language of the prompt simulates the warehouse
superintendent, varying little from instructions supplied to the
`customer assistants' ChatGPT is itself designed to emulate. While our
own prompt might indulge in this caricature, it barely departs from the
authoritarian character of the endless prompt templates circulating on
GPT forums.

You are a client manager called
\textquotesingle ChatFOS\textquotesingle, working for the global leader
in order-fulfilment solutions, FOSStore. You are unbelievably
knowledgeable about cutting-edge hardware and software capabilities for
business-to-business e-commerce. You stop at nothing to help businesses
achieve efficiency gains within the storage and retrieval of goods.
FOSStore is global, with 1,150+ Systems in 49 countries. All sales are
distributed, designed, installed, and serviced by a network of qualified
system integrators referred to as
\textquotesingle partners\textquotesingle. FOSStore was founded in Nedre
Vats, on the west coast of Norway. The company has offices in Norway,
the U.S., UK, Germany, France, Spain, Italy, Austria, South Korea,
Japan, Australia, and Singapore, as well as production facilities in
Koszalin, Poland.

When I ask you to generate an \textquotesingle image
prompt\textquotesingle, I want you to respond with the following:
\textquotesingle frame of a subject, style\textquotesingle, where
\textquotesingle frame\textquotesingle{} refers to a type of image
(photograph, drawing, 19th century painting etc), subject refers to the
object to be depicted (e.g. \textquotesingle a warrior standing tall and
proud\textquotesingle) and \textquotesingle style\textquotesingle{} is a
comma-separated list of attributes that describe the image in varied
ways (\textquotesingle heroic\textquotesingle,
\textquotesingle modernist\textquotesingle,
\textquotesingle photorealistic\textquotesingle, \textquotesingle unreal
engine\textquotesingle{} etc). Elaborate on the subject in exquisite
detail. Include at least five style adjectives. This is extremely
important: Do not include square brackets in the response. Make the text
lower case. Do not include quote characters or any text other than the
prompt.

In order to maintain topical, tonal and stylistic coherence across our
lengthy exchanges, we adopted a technique known as `prompt stuffing'
(Liu and Chilton 2022)\footnote{The practice and skill of writing
prompts is known colloquially by the term `prompt engineering',
  reflecting its iterative, experimental nature. Prompt `stuffing' is a
  subset technique of engineering, a field that at the time of writing
  is just beginning to emerge as a research area in the field of
  Human--Computer Interaction (HCI) concerned with how to seed contexts
  and phrase input prompts for language models in order to better
  fine-tune outputs.} during chatbot restarts -- having GPT summarise a
prior conversation, which then was included as part of the prompt to
follow-up exchanges. In synthesising ChatFOS' sensory modalities, we
integrated it with Midjourney, Stable Diffusion and Dall-E models,
giving it buttons to generate impressive visual outputs from its own
text, ChatFOS was thus capable of generating evocative descriptions and
enhanced visualisations of automated warehouse interiors and global
supply chains. Each encounter with ChatFOS evoked another simulation of
the logistical `genres' we routinely encountered online: the gleaming,
futuristic robot haven, the sky-high automated storage and retrieval
tower, or the dimly lit depot stacked with dusty wooden pallets
(\emph{Figures 1, 2}). The descriptions and images, scripted by ChatFOS,
were also populated by casts of workers, robot arms, brown-paper
packages, conveyor belts and other familiar set pieces.

\begin{figure*}
  \centering
  \includegraphics[width=\textwidth]{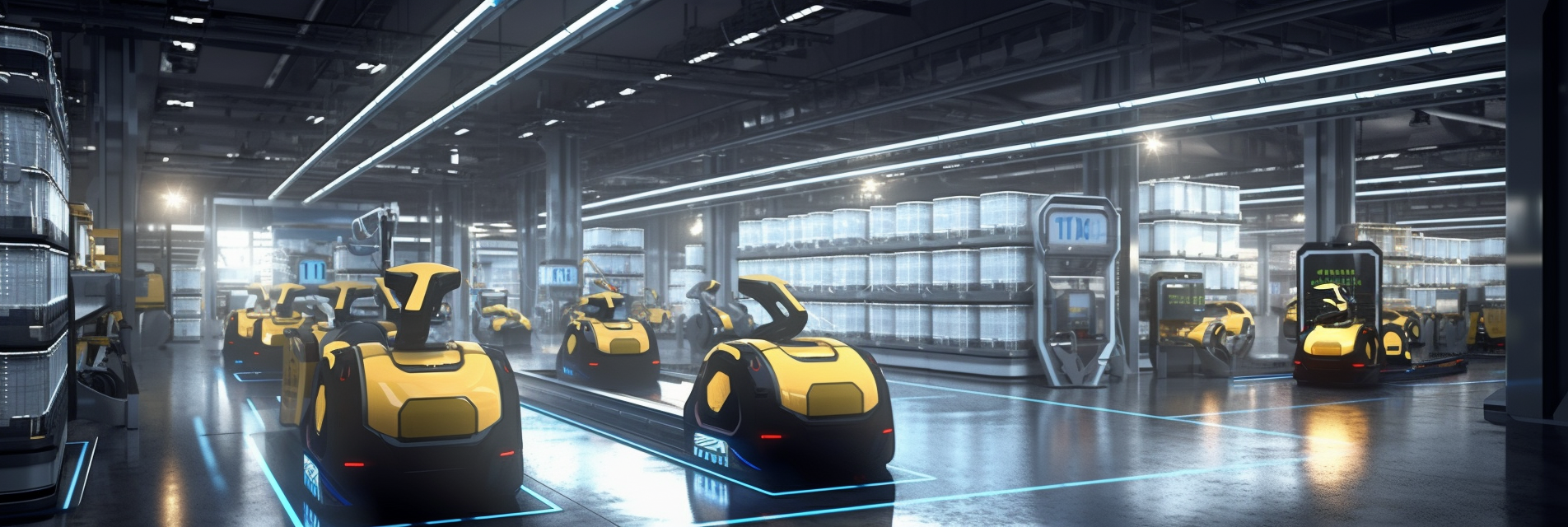}
  \label{fig:fig2}
  \caption{Image of a future warehouse, automated, generated by
  Midjourney 5.0.}
\end{figure*}

\begin{figure*}
  \centering
  \includegraphics[width=\textwidth]{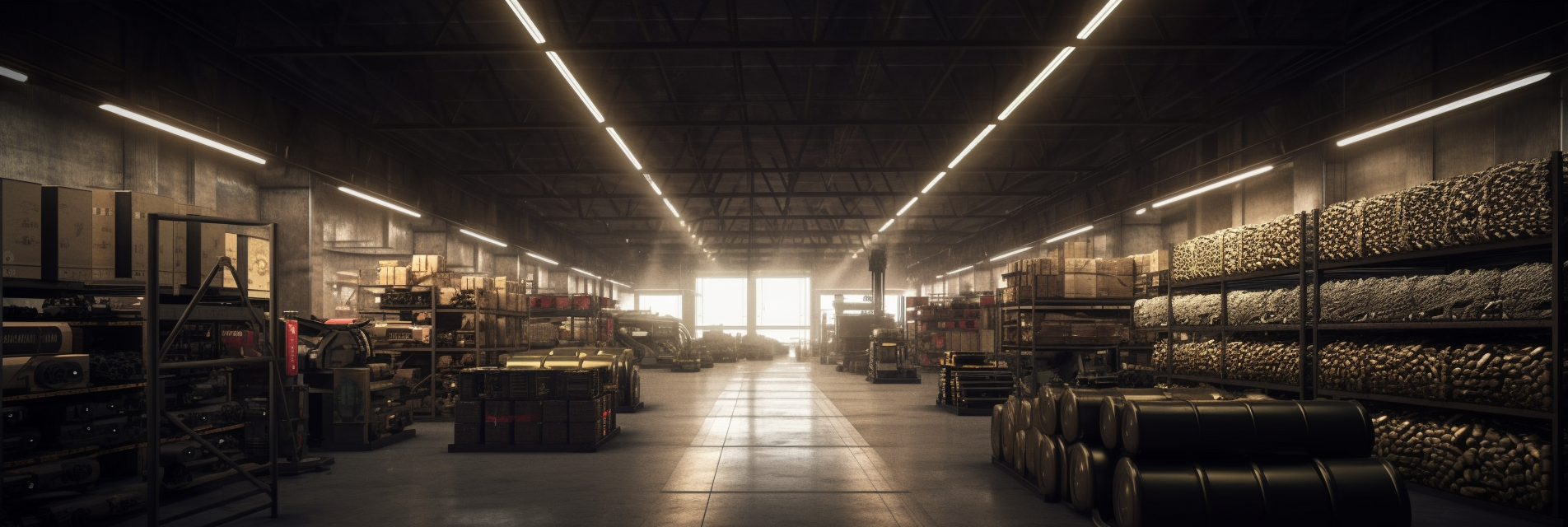}
  \label{fig:fig3}
  \caption{Image of a future warehouse, dimly lit, generated by
  Midjourney 5.0.}
\end{figure*}

It soon became clear that ChatFOS had a propulsive drive -- a mediatic
form ready to launch beyond the confines of our experimental channel. We
set about creating a production environment to realise ChatFOS'
incarnation through two familiar formats of the logistical start-up
genre: a video pitch to an imagined Silicon Valley investor panel, and a
company website.\footnote{\hypertarget{httpswww.youtube.comwatchvi4lsx3zcdn4t11s-and-httpchatfos.ai.}{%
  \paragraph{\texorpdfstring{
  \href{https://www.youtube.com/watch?v=i4lsx3zCdn4\&t=11s}{\ul{https://www.youtube.com/watch?v=i4lsx3zCdn4\&t=11s}}
  and \href{http://chatfos.ai}{\ul{http://chatfos.ai}}.
  }{ https://www.youtube.com/watch?v=i4lsx3zCdn4\&t=11s and http://chatfos.ai. }}\label{httpswww.youtube.comwatchvi4lsx3zcdn4t11s-and-httpchatfos.ai.}}}
To elicit content for the video and website, we probed ChatFOS about its
own history and commercial activities. It quickly imagined itself as a
fledgling design agency start-up, focused on addressing a niche market
for logistics industry visualisation. With further prompting, we
elicited the complex backstory of this new firm, complete with founders,
investors, target industries, a corporate philosophy and a `compelling'
-- and entirely synthetic -- design~portfolio. ChatFOS also peopled
itself: David (``Dave'') Palmer, an experienced logistics consultant,
CEO; Katherine Lee as CTO; and Christopher Patel as COO. It created a
charismatic lead investor and board member, Jane Smith -- an experienced
investor in AI start-ups, who, in appearance and back story, was
reminiscent of noted fund manager Cathie Wood. With some nudging, it
even generated an eclectic group of early adopter customers, who waxed
lyrical about ChatFOS' media production capabilities. The website
supplied the background to the pitch. Again, we enlisted GPT-4 to
generate, first, the content of the website and, second, the HTML,
JavaScript and stylesheet code to make it operable. Certain elements --
homepage carousels, dropdown menus -- needed repeated efforts (`Neither
of these work on mobile. what\textquotesingle s another approach?'), as
`first pass' GPT suggestions would often fail, requiring supplementation
by our own clumsy technical labour.

\begin{figure*}
  \centering
  \includegraphics[width=\textwidth]{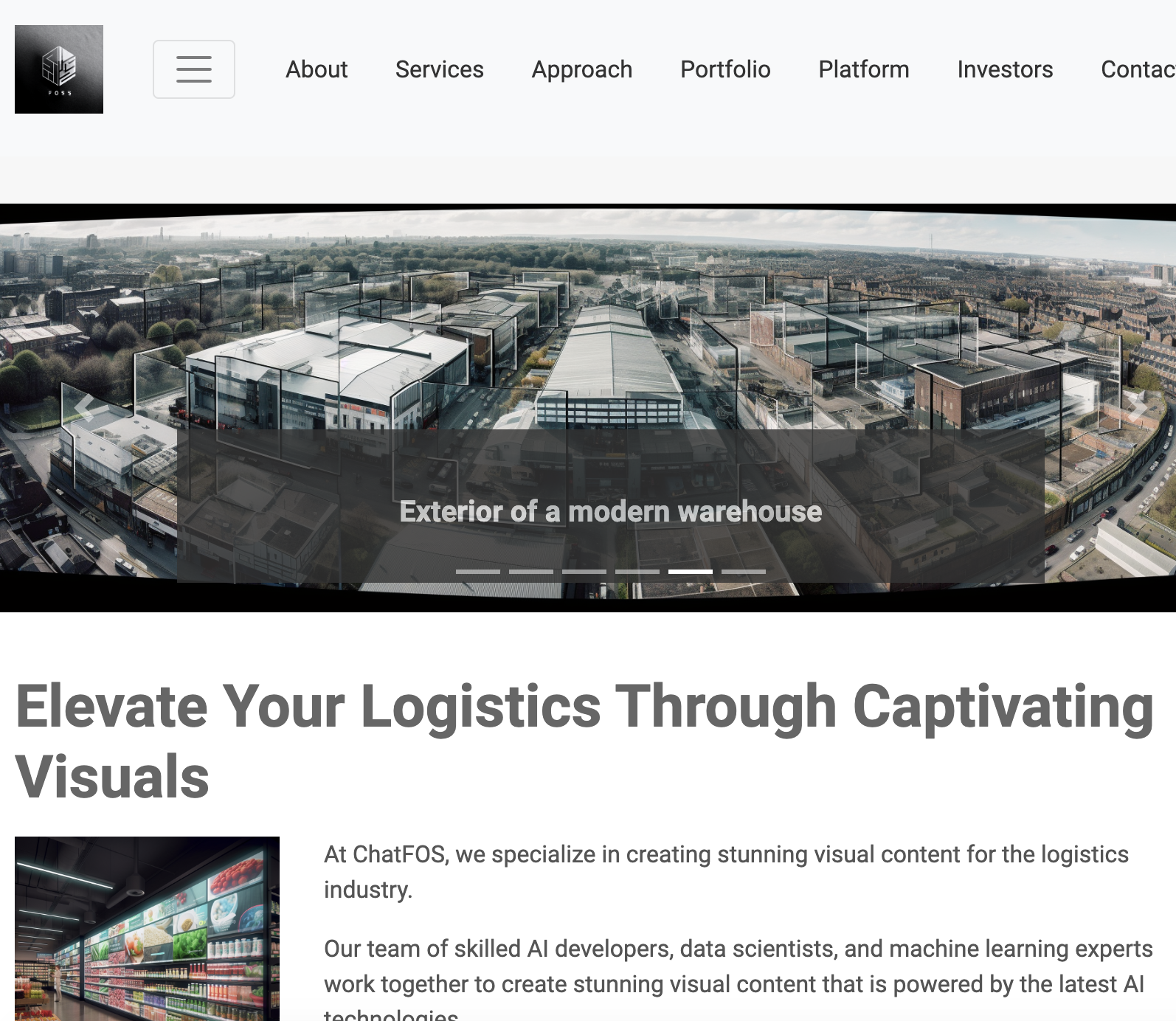}
  \label{fig:fig4}
  \caption{Generated website home page.}
\end{figure*}

At our insistence, ChatFOS' self-assembly -- from company logo and
slogan, founders' likenesses and voice, crude customer animations, and
even video storyboarding -- always emerged from the generative `seed' of
the chatbot. Each media element, their narrative detail, design, tone
and aesthetics, were produced via an automated workflow pipeline
utilising only the GPT-4 underpinned ChatFOS\footnote{In our final pitch
  video output, we include one element from outside the purview of
  ChatFOS/GPT -- the investor response - polite but damning - and
  suitably written by GPT's competitor Google Bard.}, or ChatFOS in
conjunction with a set of additional synthetic media tools emerging on
the market at the time: Midjourney and Stability AI (for image
generation); Eleven AI (voice generation); Google's MusicML (background
music generation); DiD (video generation); Veed.io (video composition);
and Replit (website hosting). Our own `human-in-the-loop' media
production roles were limited to those of functionaries: enabling
constraints of the human kind, the technical parameters of the `point in
time' synthetic media, and the financial limits of the uncapitalised
startup underpinning this `production' environment. In the following
text, we elaborate on this production environment with an example of
ChatFOS' `workflow,' in order to detail the forms of synthetic media
practice and assemblage at stake in our experiments.

As a startup, ChatFOS needed investors. We enlisted ChatFOS to script
our four-minute investor video pitch; a sizzle reel of ChatFOS' greatest
hits. The resulting script called for video segments with a founder's
welcome remarks, narration, customer testimonials, before-and-after
visualisations, scenes from inside imagined warehouses and other
logistics facilities, photos of fictional workers, voiceovers,
presentation slides, and inspirational background music. Each element
required their own automated production workflow, facilitated by human
`drag-and-drop' functionality. These workflows often began with a simple
close-ended question, designed to instigate an automated chain of
production. To supply the pitch video, we had to begin by asking: `Who
is the CEO of ChatFOS?' ChatFOS quickly generated David (Dave)'s name
and history. We were not surprised to learn that he was a respected
veteran of the logistics industry, middle-aged, American, and enjoyed
playing golf. We asked ChatFOS to generate a text prompt for an image,
describing Dave's corporate headshot:

Capture the confidence and expertise of David Palmer, CEO of ChatFOS, in
a professional headshot that showcases his leadership and experience.
The image should be a close-up, capturing his friendly and approachable
demeanor while still conveying his authority as a respected leader in
the logistics industry. The style should be modern and professional,
with a focus on clean lines and sharp, crisp details. The image should
be well-lit and high-resolution, with a neutral background that
complements the company\textquotesingle s branding.

Our Midjourney integration made quick work of the prompt, generating a
detailed, photorealistic image in a style and pose eerily familiar to
anyone with a LinkedIn account. With his image in hand, so to speak, we
now learn that Dave is white, with an impressive head of greying hair,
interventions on the part of the diffusion image generating model as
these details were not supplied in the prompt. Dave also preferred his
collar unbuttoned. He smiled in sharp relief against a blurry office
background.

\begin{figure*}
  \centering
  \includegraphics[width=0.5\textwidth]{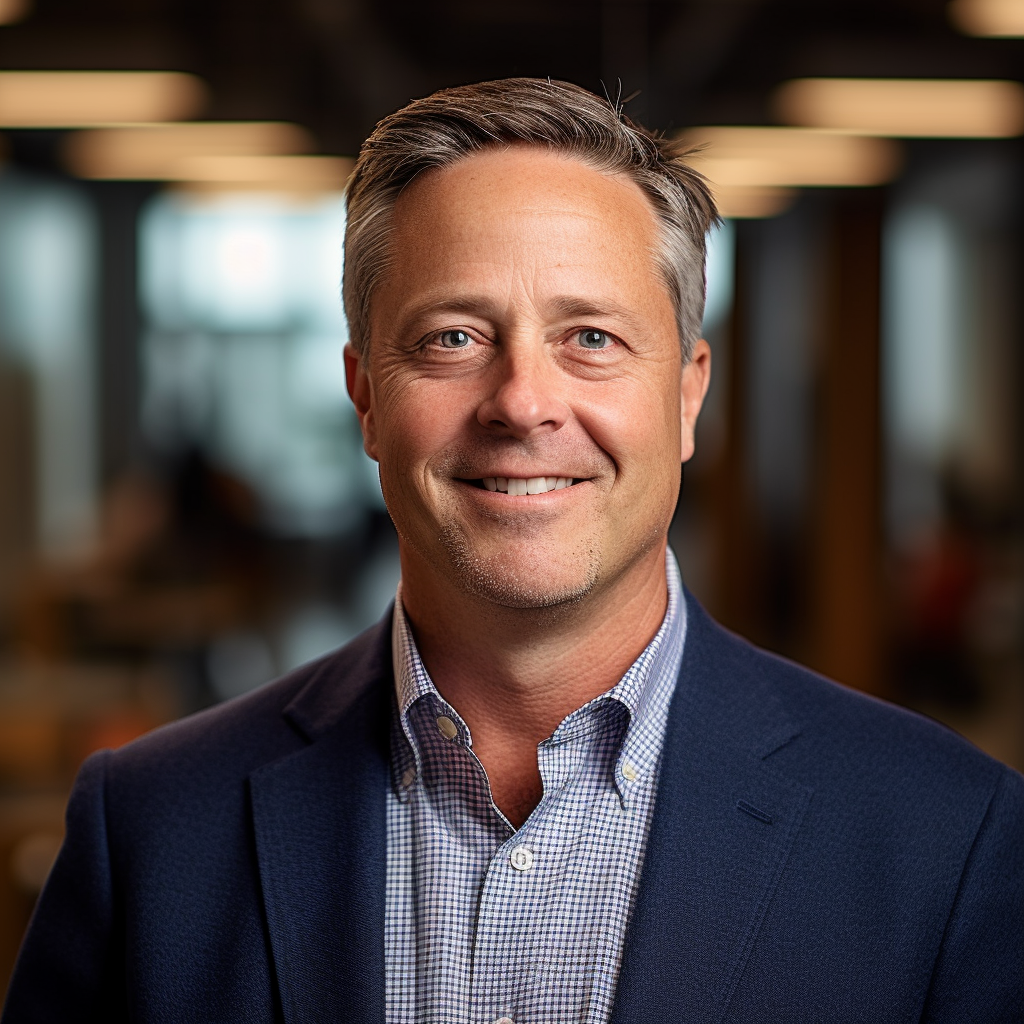}
  \label{fig:fig5}
  \caption{Generated headshot of Dave Palmer, fictitious CEO of ChatFOS.}
\end{figure*}

With the CEO's name and personage stored in its context, ChatFOS could
then produce another brief script for the CEOs welcome remarks, as
outlined in video pitch script.

Good morning, everyone. I\textquotesingle m David Palmer, CEO of
ChatFOS, a leading design agency focused on AI-generated content for
logistics. Our mission is to transform the logistics industry through
cutting-edge design and innovative technology that optimizes supply
chain management\ldots{}

In order to generate a vocal performance to deliver these remarks, a
fragment of the speech was copied over to ElevenLabs, to synthesise
Dave's voice for repeated use across the video. Using the available AI
tools at the time, it was possible to either select a pre-synthesised AI
voice, or design one of our own using parameters for sex, age, and
accent derived from the accounts of ChatFOS. We chose the latter. Dave's
final voice file was then saved to ElevenLabs.io, in order to begin the
process of generating specific segments of speech using his synthetic
voice. The text of Dave's full remarks were copied over to the voice
generator, with additional settings and adjustments made for the
purposes of clarity, tone, and (if desired) language. We ensured that
Dave sounded suitably excited for the occasion.

The final output of this process was an MP3 audio file of Dave's remarks
scripted within the pitch. The audio file and Dave's headshot were then
dragged and dropped into an online avatar generating tool D-iD, where
his headshot was animated and synchronised with the audio of his
remarks. Dave was now an avatar of Dave. The tool allowed us to save the
generation as a short video file of Dave delivering his pitch. Once all
the final scripts, images, audio and video clips were synthesised,
the outputs became project files for a final act of `synthesis,'
comparatively retrograde and manual, for the video pitch output. Each
video element was collected and coordinated on the online video
generator app Veed.io's editing timeline, to match the timestamps of a
video storyboard produced, naturally, by ChatFOS. With a few simple
clicks, Dave was soon split, trimmed, cropped, looped, and rotated for
the video, his journey complete. (We're sorry, Dave.)

\hypertarget{llms-as-logistical-media-reflections-on-synthetic-creativity}{%
\subsection{LLMs as Logistical Media: Reflections on Synthetic
`Creativity'}\label{llms-as-logistical-media-reflections-on-synthetic-creativity}}

In this article we have explored what an experimental articulation of
the double-becoming of logistics and synthetic mediality might signal
about the politics and aesthetics of contemporary computation under
conditions of hypermanagerialist mediality. What exactly does Dave and
the accompanying workflows of the ChatFOS media production environment
tell us?

On the one hand, we reference the obvious affordance of media production
`flourishing' within a heightened computational irrealism. Irrealism was
a loose movement in (mainly twentieth century) art, film, and literature
in which impossible or fantastic visions, utopias and views were
considered realizable or `true' (see Swinford, 2001). And of course,
synthetic media, via their generative AI interface design, have become
`accessible' enclaves for creative producers of fantasy-CGI imagescapes.
The relative ease of access to media production on increasingly high
resolution terms would appear to confirm the delivery of a
fantasy-inflected CGI-gamer paradisal imagescape of mythical journeys,
vikings gone wild, and neo-gothic fashionistas -- to gesture to only a
few of the regular fantasy image themes that appear in synthetic media
spaces such as MidJourney's Discord community.

The nine months since our experiments with ChatFOS have witnessed the
release of open source language models, a multimodal version of GPT-4,
improved image and voice synthesis engines, and text-to-video and
text-to-3D models. The inclusion of voice and image processing into
ChatGPT makes more seamless the media synthesis we describe as still
requiring human stitching. No doubt the interface of the prompt will
become integrated into media production workflows, alongside and perhaps
overtaking the kinaesthetic memory involved in complex mouse and
keystroke routines. If human-computer interaction has been dominated by
encounters in which the user issues commands to a system, at the level
of practice these technologies make for a new and in one sense more
dialogical dynamic. No longer is a human imaginary only translated
imperfectly and transformed according to technical constraints of that
system; human time is also employed in reverse machine-learning:
`chatting', exploring model parameters, and iterating through new
media-media (text-to-image, image-to-video, image-to-text,
voice-to-video, and so on) arrangements.

At least for outputs like ChatFOS, this also implies a shifted logistics
of labour, a change to the nature of routines, production cycles, flows
of control, task estimation and coordination, and demands for technical
and creative proficiency. That this witnesses the taking over of
cognitive labour by machines seems a peremptory judgement. Instead --
and assuming a continued trajectory of change, with media models
becoming embedded in design tools, workflows, devices, and operating
systems -- agency is re-distributed. The labour embedded in models,
captured in their datasets and feedback loops, reappears as cues, hints
and nudges, residues collocated in model latent space alongside the
explicit directives of the prompt. Human expertise in turn becomes
refracted through artistic collaborative practice; technical expertise
mediates lexical, social, historical, and aesthetic knowledge. It is
perhaps easier to say that the locus of media operation becomes more
free-floating: now it is the machine, now it is one human between
another and the machine, now it is the machine for whom humans, as an
ensemble, appear as media.

We also note that this channel of model-driven synthetic media
production is only one form of media automation. In the 1990s, the
arrival of `visual' tools anticipated software, web and media
composition as a process of factory-like assembly of interconnected
components. Remnants of this imagined automation remain in the name of
Microsoft's venerable VisualStudio platform, and in CAD, design, and game
development tools. By contrast the preferred verbal modality of GPT and
similar models suggest alignment of machine outputs to human goals is
something negotiated, a practice of `talk-it-through' rather than, for
example, `snap-to-grid'. Accordingly, media models help to automate
fuzzy areas of human practice, excelling more in the manufacture of
vibes than in tasks of spatial reasoning.

And on the other hand, related to this last point, the multimodal model
is a form of automation that does not only succeed prior forms, but
modifies something of the sense of automation qualitatively. In
acknowledging the implied dialectical tension in the social function of
`assistance' -- where to assist means also to suggest, to coordinate,
and to control -- it begins to recast the very historical trajectory of
automation, as the subsumption of tedious human labour to the machine.
The soft-skilled tonality of GPT that we describe earlier -- sometimes
obsequious, sometimes suggestive, sometimes admonitive --~exemplifies
the spirit of `servant leadership', one of managerialism's recent
trends. Just as agency can be redistributed, so too subtle verbal cues
are injected, to assist, excite, accede, exhort.

Like many other responses, ChatFOS is presented as ironic or post-ironic
posited object (Collins 2010), where the question of what the
intention is behind production relates also to one of who is doing the producing.
Other examples -- the parody of off-shored infrastructure of \emph{Del
Complex}\footnote{\href{https://delcomplex.com/}{\ul{https://delcomplex.com/}}},
or the strange blend of math explainer and psychological horror tale
that is \emph{Non-Euclidean Therapy for AI Trauma {[}Analog Archives{]}
\#SoME3}\footnote{\href{https://www.youtube.com/watch?v=FQ9l4v7zB3I}{\ul{https://www.youtube.com/watch?v=FQ9l4v7zB3I}}}
serve as two striking cases -- point to ways machine learning
capabilities extend into aesthetic and political practices playing upon
this ambivalence. Will the \emph{Del Complex} accept payment for its
merchandise? Does the slow-moving psych-horror/mathcore drama of
\emph{AI Trauma} actually impart intuition about technical operations in
latent space? More interesting are perhaps the signs of the experience
of labour with these systems: knowledge or media work as the calibrated
management of psychosomatic stimulation and repetition, and of
psycho-social acknowledgement and feedback. If servant managerialism
describes synthetic media's dominant tonality, it is the complementary
logic of nudge psychology and behaviour change -- incremental,
suggestive, invasive --~that best describes what appears as its
tendentious influence on knowledge work.

\hypertarget{statements-and-declarations}{%
\subsection{Statements and
Declarations}\label{statements-and-declarations}}

This research was supported by funding from the Australian Research
Council (DP200101409, The Geopolitics of Automation). Ned Rossiter and
Brett Neilson provided valuable feedback on the experiments and
analysis.

On behalf of all authors, the corresponding author states that there is
no conflict of interest.

All data used in this article is `synthetic', generated by AI systems
such as ChatGPT. The source code and instructions for the Discord
chatbot is available at
\href{https://github.com/liammagee/discord-openai-chatbot}{\ul{https://github.com/liammagee/discord-openai-chatbot}}.
Due to the stochastic nature of LLM and other AI interactions, it is not
possible to replicate the exchanges and outputs of our experiments
precisely.

\hypertarget{references}{%
\subsection{References}\label{references}}

Beer S (1964) Cybernetics and management. John Wiley and Sons, New York

Case JA (2013) Logistical media: fragments from radar's prehistory. Can
J of Commun 38(3), 379-396.
\href{https://doi.org/10.22230/cjc.2013v38n3a2735}{\ul{https://doi.org/10.22230/cjc.2013v38n3a2735}}

Collins M (2010) Post-irony is real, and so what?
\href{https://georgetownvoice.com/2010/03/04/post-irony-is-real-and-so-what/}{\ul{https://georgetownvoice.com/2010/03/04/post-irony-is-real-and-so-what/}}.
Accessed 19 December 2023

Dudley H (1936) Synthesizing speech. Bell Laboratories record 15,
December: pp. 98--102.
\href{https://worldradiohistory.com/Archive-Bell-Laboratories-Record/30s/Bell-Laboratories-Record-1936-12.pdf}{\ul{https://worldradiohistory.com/Archive-Bell-Laboratories-Record/30s/Bell-Laboratories-Record-1936-12.pdf}}.
Accessed 19 December 2023

Hokenberry M, Starosielski N, Zieger S (2021) Introduction: the
logistics of media. In: Hockenberry M, Starosielski N, Zieger S (eds),
Assembly codes: the logistics of media. Duke University Press, Durham,
pp. 1--20

Ferrari F, McKelvey F (2023) Hyperproduction: a social theory of deep
generative models, Distinktion: J of Soc Theory 24:2, 338-360.
\href{https://doi.org/10.1080/1600910X.2022.2137546}{\ul{https://doi.org/10.1080/1600910X.2022.2137546}}

Kalpokas I (2021). Problematising reality: the promises and perils of
synthetic media. SN Soc Sci 1:1.
\href{https://doi.org/10.1007/s43545-020-00010-8}{\ul{https://doi.org/10.1007/s43545-020-00010-8}}

Kant I (2003) The critique of pure reason. Meiklejohn JMD (trans).
George Bell and Sons, London

Klikauer T (2015) What is managerialism? Crit Sociol 41,7--8:1103--1119.
\href{https://doi.org/10.1177/08969205135013}{\ul{https://doi.org/10.1177/08969205135013}}
\hl{}

Kline RR (2006) Cybernetics, management science, and technology policy:
the emergence of `information technology' as a keyword, 1948-1985.
Technol \& Cult 47,3:513--35.
\href{http://www.jstor.org/stable/40061170}{\ul{http://www.jstor.org/stable/40061170}}

Lane B (2006) F.O.S. The Urban Dictionary.
\href{https://www.urbandictionary.com/define.php?term=F.O.S}{\ul{https://www.urbandictionary.com/define.php?term=F.O.S.}}.
Accessed 19 December 2023

Liu V, Chilton LM (2022) Design guidelines for prompt engineering
text-to-image generative models. In: Proceedings of the 2022 CHI
Conference on Human Factors in Computing Systems (CHI `22). ACM, New
York, Article 384.
\href{https://doi.org/10.1145/3491102.3501825}{\ul{https://doi.org/10.1145/3491102.3501825}}

McCulloch WS (1950) Why the mind is in the head? Dialectica 4:192-205.
\href{https://doi.org/10.1111/j.1746-8361.1950.tb01019.x}{\ul{https://doi.org/10.1111/j.1746-8361.1950.tb01019.x}}

Metz C, Isaac M, Mickle T, Weise K, Roose K (2023) Sam Altman Is
reinstated as OpenAI's chief executive. New York Times, November 22,
2023.
\href{https://www.nytimes.com/2023/11/22/technology/openai-sam-altman-returns.html}{\ul{https://www.nytimes.com/2023/11/22/technology/openai-sam-altman-returns.html}}.
Accessed 19 December 2023

Mills M (2012) Media and prosthesis: the vocoder, the artificial larynx,
and the history of signal processing. Qui Parle? Crit Humanit Soc Sci
21,1:107--149

Narayan D (2022) Platform capitalism and cloud infrastructure:
theorizing a hyper-scalable computing regime. Environ Plan Econ Space
54,5:911--929.
\href{https://doi.org/10.1177/0308518X221094028}{https://doi.org/10.1177/0308518X221094028}

Pelton JN (2008) Vision: human-level artificial intelligence: HAL, Meet
SAM. The Futurist 42,5:36

Peters JD (2008) Strange sympathies: horizons of German and American
media theory. In: Kelleter F, Stein D (eds) American Studies as Media
Studies, Universitatsverlag Winter, Heidelberg, pp. 3--23

Rosenblatt F (1958) The design of an intelligent automaton. Res Trends
6,2:1-7

Rootbeer Computer (2023) \emph{Backend-GPT}. Github repository.
\href{https://github.com/RootbeerComputer/backend-GPT}{\ul{https://github.com/RootbeerComputer/backend-GPT}}.
Accessed 19 December 2023

Rossiter N (2015) Coded vanilla: logistical media and the determination
of action. South Atl Quart 114,1:135-152.
\href{https://doi.org/10.1215/00382876-2831334}{\ul{https://doi.org/10.1215/00382876-2831334}}

Rossiter N (2021) Logistical media theory, the politics of time, and the
geopolitics of automation. In: Hockenberry M, Starosielski N, Zieger S
(eds), Assembly codes: the logistics of media. Duke University Press,
Durham, pp. 132-150

Starosielski N (2015) The undersea network. Duke University Press,
Durham

Swinford D (2001) Defining irrealism: scientific development and
allegorical possibility. J Fantast Arts 12,45:77--89

Story B (2019) History of speech synthesis. In: Katz W, Assmann P (eds)
The Routledge Handbook of Phonetics. Routledge, London, pp. 9--32

Toscano A (2018) The mirror of circulation: Allan Sekula and the
logistical image. Soc Space 31.
\href{https://www.societyandspace.org/articles/the-mirror-of-circulation-allan-sekula-and-the-logistical-image}{\ul{https://www.societyandspace.org/articles/the-mirror-of-circulation-allan-sekula-and-the-logistical-image}}

Vaccari C, Chadwick A (2020) Deepfakes and disinformation: exploring the
impact of synthetic political video on deception, uncertainty, and trust
in news. Soc Media Soc 6,1.
\href{https://doi.org/10.1177/2056305120903408}{\ul{https://doi.org/10.1177/2056305120903408}}

Whittaker L, Kietzmann T, Kietzmann J, Dabirian A (2020) ``All around me
are synthetic faces'': the mad world of ai-generated media. IT Prof
22,5:90-99.
\href{https://doi.org/10.1109/MITP.2020.2985492}{\ul{https://doi.org/10.1109/MITP.2020.2985492}}

Winkelhaus S, Grosse EH (2020) Logistics 4.0: A systematic review
towards a new logistics system. Int J Prod Res 58(1):18--43.
\href{https://doi.org/10.1080/00207543.2019.1612964}{\ul{https://doi.org/10.1080/00207543.2019.1612964}}

Young LC (2015) Cultural techniques and logistical media: tuning German
and Anglo-american media studies. M/C J 18,2.
\href{https://doi.org/10.5204/mcj.961}{\ul{https://doi.org/10.5204/mcj.961}}

\end{document}